\documentclass[prb,twocolumn,amsmath,amssymb,floatfix,superscriptaddress]{revtex4-2}
\usepackage{graphicx} 
\usepackage{booktabs}
\usepackage[dvipsnames]{xcolor}
\usepackage[version=4]{mhchem}
\usepackage{xspace, bm, braket}
\usepackage{siunitx}
\usepackage{verbatim}
\usepackage{subfigure}
\usepackage[colorlinks, citecolor=blue, linkcolor=blue, urlcolor=blue, breaklinks]{hyperref}
\usepackage{hyperref}
\usepackage[normalem]{ulem}

\newcommand{\PNO}{\ce{Pr4Ni3O8}}
\newcommand{\NNO}{\ce{NdNiO2}}
\newcommand{\CCO}{\ce{CaCuO2}}

\newcommand{\imag}{\text{i}}
\newcommand{\angstrom}{\textup{\AA}}

\begin{document}

\title{Comparative Many-Body Study of \PNO{} and \NNO{}}

\author{Jonathan Karp}
\email{jk3986@columbia.edu}
\affiliation{Department of Applied Physics and Applied Math, Columbia University, New York, NY 10027, USA}

\author{Alexander Hampel}
\affiliation{Center for Computational Quantum Physics, Flatiron Institute, 162 5th Avenue, New York, NY 10010, USA}

\author{Manuel Zingl}
\affiliation{Center for Computational Quantum Physics, Flatiron Institute, 162 5th Avenue, New York, NY 10010, USA}

\author{Antia S. Botana}
\affiliation{Department of Physics, Arizona State University, Tempe, AZ 85287}

\author{Hyowon Park}
\affiliation{Materials Science Division, Argonne National Laboratory, Lemont, IL 60439}
\affiliation{Department of Physics, University of Illinois at Chicago, Chicago, IL 60607}

\author{Michael R. Norman}
\affiliation{Materials Science Division, Argonne National Laboratory, Lemont, IL 60439}

\author{Andrew J. Millis}
\affiliation{Center for Computational Quantum Physics, Flatiron Institute, 162 5th Avenue, New York, NY 10010, USA}
\affiliation{Department of Physics, Columbia University, New York, NY 10027, USA}

\date{\today}

\begin{abstract}
We study the many-body electronic structure of the stoichiometric and electron-doped trilayer nickelate \PNO{} in comparison to that of the stoichiometric and hole-doped infinite layer nickelate \NNO{} within the framework of density functional plus dynamical mean field theory, noting that \PNO{} has the same nominal carrier concentration as \NNO{} doped to a level of 1/3 holes/Ni. We find that the correlated Ni-$3d$ shells of both of these low valence nickelates have similar many-body configurations with correlations dominated by the $d_{x^2-y^2}$ orbital. Additionally, when compared at the same nominal carrier concentration, the materials exhibit similar many-body electronic structures, self energies, and correlation strengths, but differ in Fermiology. Compared to cuprates, these materials are closer to the Mott-Hubbard regime due to their larger charge transfer energies.  Moreover, doping involves the charge reservoir provided by the rare earth $5d$ electrons, as opposed to cuprates where it is realized via the oxygen $2p$ electrons.
\end{abstract}

\maketitle

\section{Introduction}

Understanding the physics and chemistry underlying the extraordinary properties  of layered copper-oxide materials has been a challenge to researchers over the more than thirty years since the discovery of superconductivity in La$_{2-x}$Ba$_x$CuO$_4$~\cite{Bednorz86} and the basic questions of mechanism for superconductivity are not yet settled. One approach to this question is to identify ``cuprate analog" materials that have similar physical and nominal electronic structure but differ in local chemistry. A key feature of the cuprates is a square planar coordinated transition metal with a nominal $d^9$ valence. In this context, \citeauthor{Anisimov99} suggested that  square planar $d^9$ nickel materials such as the ``infinite layer" $R$NiO$_2$ with $R=$ La, Pr, Nd, or other rare earth elements would provide an important comparison \cite{Anisimov99}, and these and related materials were studied theoretically within various density functional (DFT and DFT+U) approximations \cite{Anisimov99, Lee04, poltavets2010bulk, pardo2010quantum, pardo2012pressure, Liu14, botana2016charge, botana2017PNO, Botana18}. Synthesis of  stoichiometric LaNiO$_2$ was reported already  in 1983 \cite{Crespin1983reduced}, followed by improvements in synthesis \cite{hayward1999sodium} and then high quality thin films \cite{kawai2009reversible, Ikeda2016direct}.  This went hand in hand with experimental studies of multilayer reduced  Ruddlesden-Popper variants \cite{poltavets2007crystal, poltavets2010bulk, cheng2012pressure, Zhang2016stacked, Zhang2017, zhang2019spin, huangfu2020short}. The reduced trilayer Ruddlesden-Popper materials were found to exhibit long-ranged (La$_4$Ni$_3$O$_8$, \cite{Zhang2016stacked, Zhang2017, zhang2019spin, lin2020strong})  or short-ranged (\PNO{}, \cite{huangfu2020short, lin2020strong}) density wave order.   

Things took a dramatic turn in 2019 when superconductivity was found upon hole doping the infinite layer material  \NNO{} \cite{Hwang2019} and subsequently PrNiO$_2$ \cite{Osada2020}. Further investigation of the phase diagram of \NNO{} has shown that it becomes superconducting on modest (13.5-22.5\%) Sr doping \cite{zeng2020phase, li2020superconducting}. However,  superconductivity has not yet been observed in the reduced Ruddlesden-Popper nickelates, nor has density wave order been found in the infinite layer materials, although in the cuprate family trilayer materials exhibit superconductivity with among the highest reported transition temperatures.  Moreover, the transport properties of 30\% hole doped \NNO{} \cite{zeng2020phase} and undoped \PNO{} \cite{Zhang2017} differ in that the former exhibits a weak localization upturn at low temperatures, whereas the latter has a more metallic behavior of the resistivity  in temperature, similar to that of overdoped cuprates. 

Several recent theoretical studies of the infinite layer materials have highlighted the possible  importance of the high-spin $d^8$ configuration \cite{kang2020infinitelayer,werner2019nickelate,lechermann2019late, wang2020hunds, petocchi2020normal, zhang2020type, wan2020calculated}.  Significant participation of $d^8$ in the infinite layer material has been suggested based on recent resonant x-ray studies \cite{hepting2019,goodge2020doping}, while related experimental studies of the trilayer material \cite{Zhang2017,lin2020strong} have found no evidence for high-spin $d^8$.  The latter studies also argued that the trilayer material is intermediate in correlation strength between the infinite layer material and the cuprates.

Motivated by these differences, in this paper, we present  a comparative density functional plus dynamical mean field theory (DFT+DMFT) \cite{Georges1996,Georges04,Kotliar06,Held06} study of the  trilayer nickelate \PNO{} and infinite layer nickelate \NNO{}. The Pr variant of the trilayer material was chosen because it exhibits metallic resistivity similar to that of overdoped cuprates \cite{Zhang2017}. The two materials exhibit some differences in three dimensional arrangement, leading to different $c$-axis dispersion and different electron count in the stoichiometic compounds. We use the virtual crystal approximation to vary the carrier concentrations so that we can compare the materials at the same doping. When compared at the same doping, we find that both have similar electronic structures, self energies, and mass enhancements, but there are significant differences in their Fermi surfaces that can be traced to their differing c-axis dispersions.  The latter will be discussed below and is related to DFT+U findings  of differences in their magnetic phase diagrams \cite{botana2017PNO}. 

The rest of this paper is organized as follows. Section ~\ref{sec:formalism} further introduces the two materials and also the formalism used here. Section ~\ref{sec:DFTresults} presents the results obtained from density functional theory calculations, with Section~\ref{sec:DMFTSelf} presenting the DFT+DMFT self energies, spectral functions, orbital occupations, and multiplet occurrence probabilities. We
offer some concluding thoughts in Section~\ref{sec:Conclusion}.

\begin{figure}[t]
    \centering
    \includegraphics[width = \columnwidth]{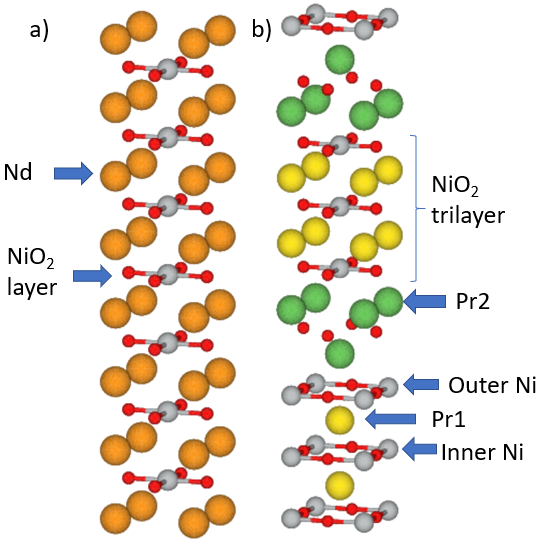}
    \caption{Left: crystal structure of infinite layer \NNO{} in the P4/mmm structure. Right: crystal structure of trilayer \PNO{} in the I4/mmm structure. Ni atoms are shown in silver, O in red, Nd in orange, Pr1 in yellow, and Pr2 in green. Crystal structures are visualized using Vesta \cite{vesta}.}
    \label{fig:structs}
\end{figure}

\section{Background and Formalism \label{sec:formalism}}
The \NNO{} crystal structure, shown in Fig.~\ref{fig:structs}, is composed of NiO$_2$ planes separated by layers of Nd. The structure of \PNO{}, also shown in Fig.~\ref{fig:structs}, is composed of blocks of three NiO$_2$ layers. The three NiO$_2$ layers in one group of three layers are separated by Pr ions, analogous to the structure of the infinite layer material, but each group of three NiO$_2$ layers is separated from the neighboring three-layer groups by a fluorite-structure Pr$_2$O$_2$ block. Each successive group is also displaced by one half of a lattice constant in the $x$ and $y$ directions so that it sits above the centers of the Ni plaquettes in the neighboring groups. This, along with the Pr$_2$O$_2$ block, means that the inter-trilayer coupling is weak enough that each three-layer group is effectively independent so that the net c-axis dispersion is much weaker than in the infinite layer material. Another effect of the Pr$_2$O$_2$ block is that it absorbs one electron from the NiO$_2$ trilayer, nominally taking $1/3$ of an electron from each Ni atom, so \PNO{} has a formal valence of $d^{8.67}$ while \NNO{} has a formal valence of $d^9$.

While there are three Ni atoms per formula unit, the two Ni atoms on the outer layers are equivalent by symmetry, so we refer to the two types as ``inner" and ``outer". There are also two types of Pr atoms, one which we call Pr1 in between individual NiO$_2$ layers but within the trilayer group, and one which we call Pr2 in between groups of three layers (i.e., in the Pr$_2$O$_2$ block). 

In this paper, we perform fully charge self consistent DFT+DMFT \cite{Georges1996,Georges04,Kotliar06,Held06} calculations using Wien2k~\cite{Blaha2018} and TRIQS \cite{TRIQS, TRIQS/CTHYB, TRIQS/DFTTOOLS}. The $4f$-states of the rare earths are treated as core electrons. We use the virtual crystal approximation at the DFT level (implemented via a fractional atomic charge on the rare earth sites) to perform the calculations for both materials at dopings corresponding to nominal Ni $d$-valences of $d^{8.67}$ and $d^9$. That is, we hole dope \NNO{} by 0.33 to compare it to undoped \PNO{}, and electron dope \PNO{} by 1 (1/3 per Ni atom) to compare it to undoped \NNO{}

We use projectors in a wide energy window of \SI{-10}{eV} to \SI{10}{eV} to capture all of the relevant Ni-$d$, O-$p$, and Pr1/Nd-$d$ states. We use a five Ni-$d$ orbital impurity model with a rotationally invariant Slater Hamiltonian with $U = \SI{7}{eV}$ and $J = \SI{0.7}{eV}$, representative of nickelates~\cite{Nowadnick15}, at a temperature $T = \SI{290}{K}$. We approximate the double counting correction using the fully localized limit (FLL) formula \cite{liechtenstein1995fll, aichhorn2011importance} (a brief discussion of alternative double counting schemes is presented in Appendix~\ref{app:dmft}). We use the single site DMFT approximation for each Ni atom and solve the impurity problem using CTHYB \cite{TRIQS/CTHYB}. In the case of \PNO{}, we have to solve two impurity problems, one for the Ni atom of the inner layer and one for the equivalent Ni atoms of the outer layers. We use the maximum entropy method to analytically continue the self energies to the real frequency axis. Further details of the calculations can be found in Appendix~\ref{app:dmft}.

\section{DFT Results \label{sec:DFTresults}}

\begin{figure*}[t]
    \centering
    \subfigure{\includegraphics[width=0.48\linewidth]{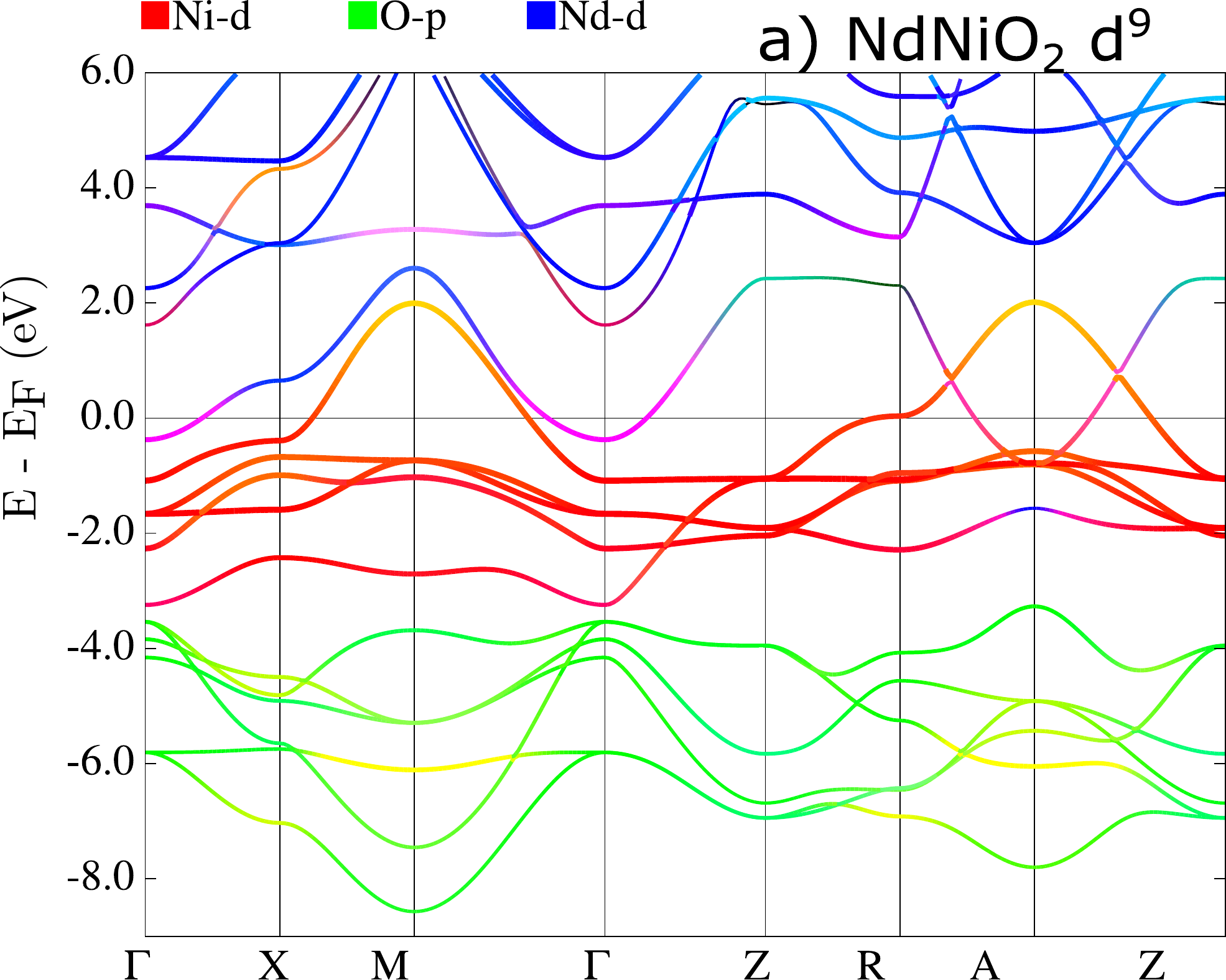}}
    \subfigure{\includegraphics[width=0.48\linewidth]{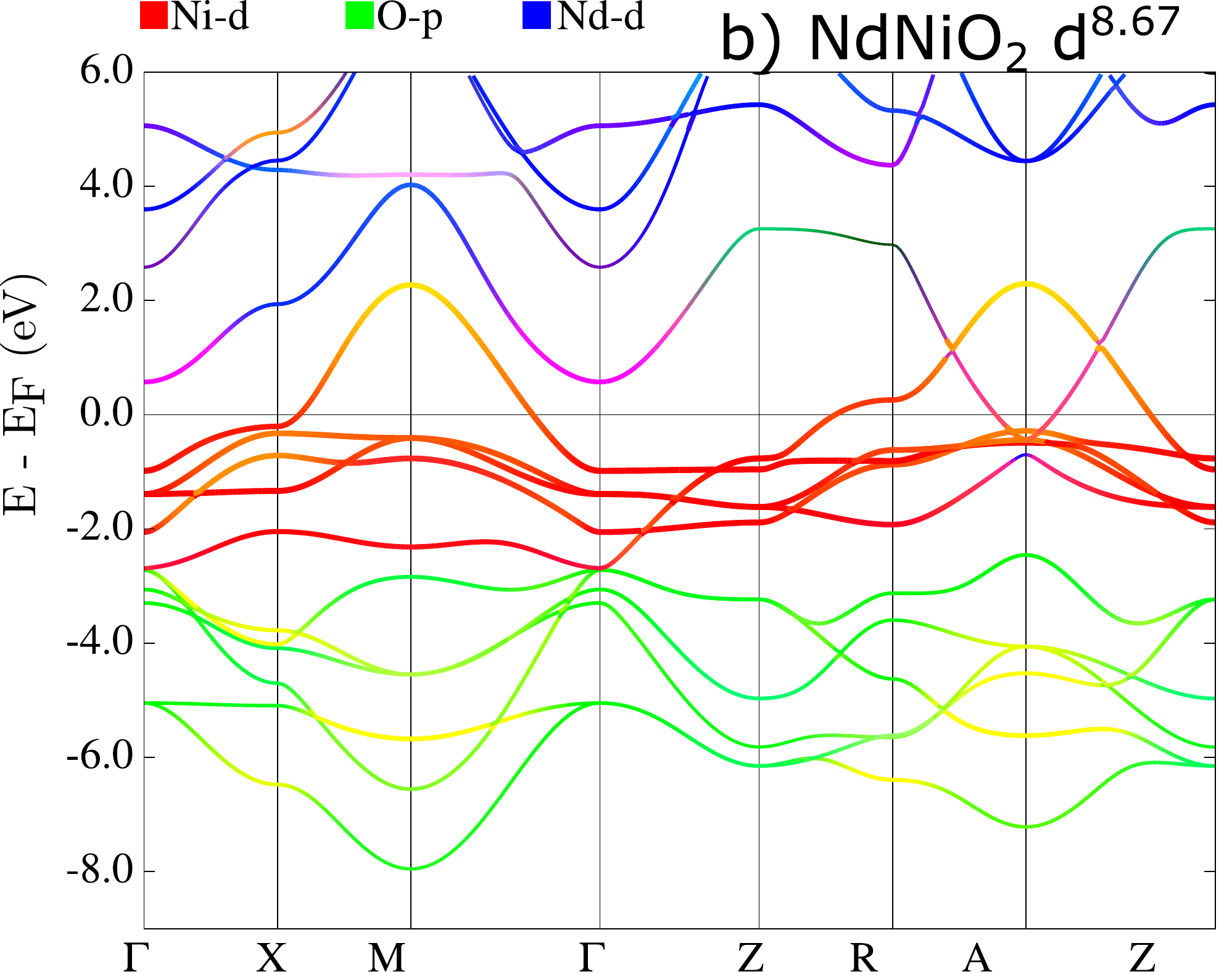}}
    \\
    \subfigure{\includegraphics[width=0.48\linewidth]{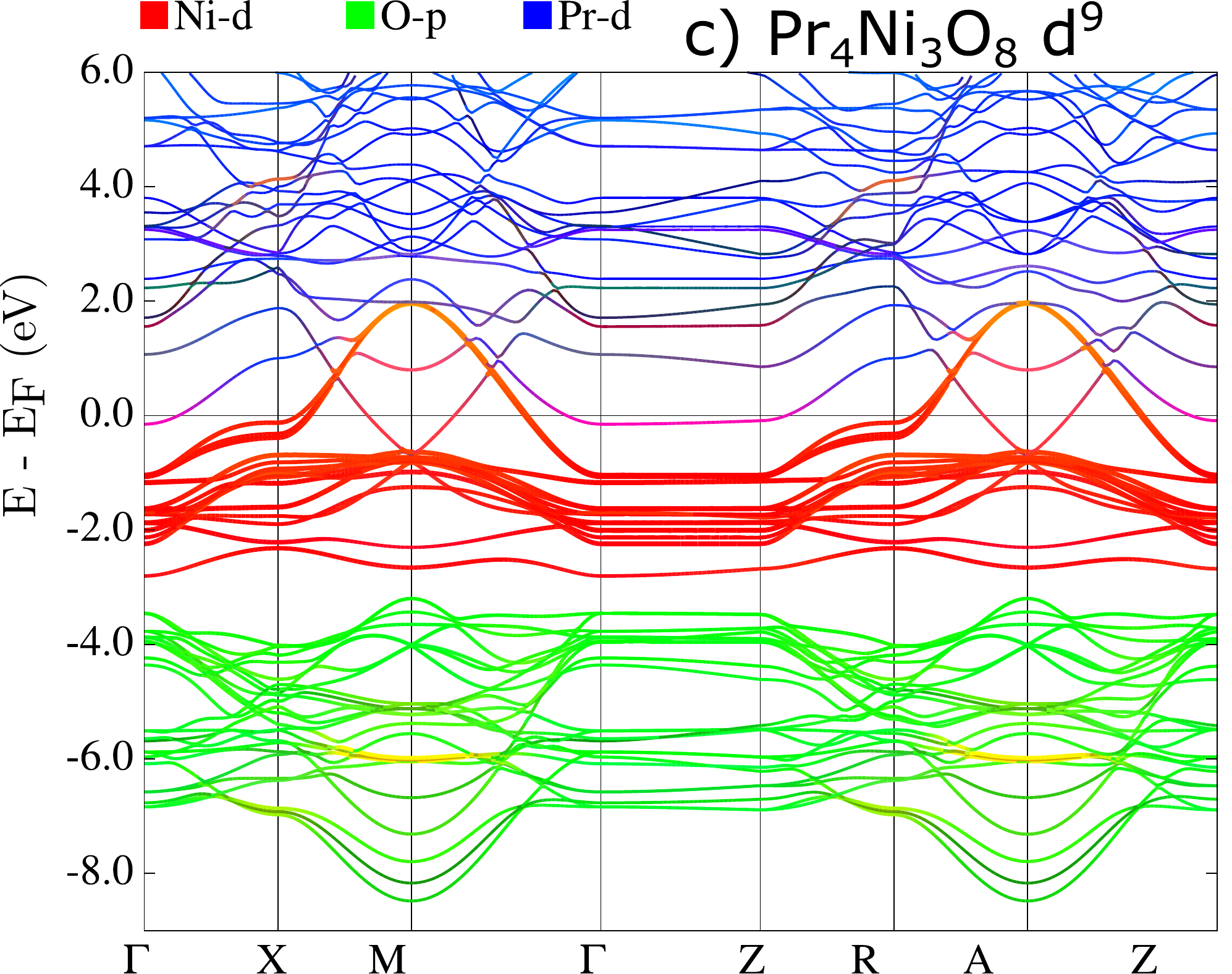}}
    \subfigure{\includegraphics[width=0.48\linewidth]{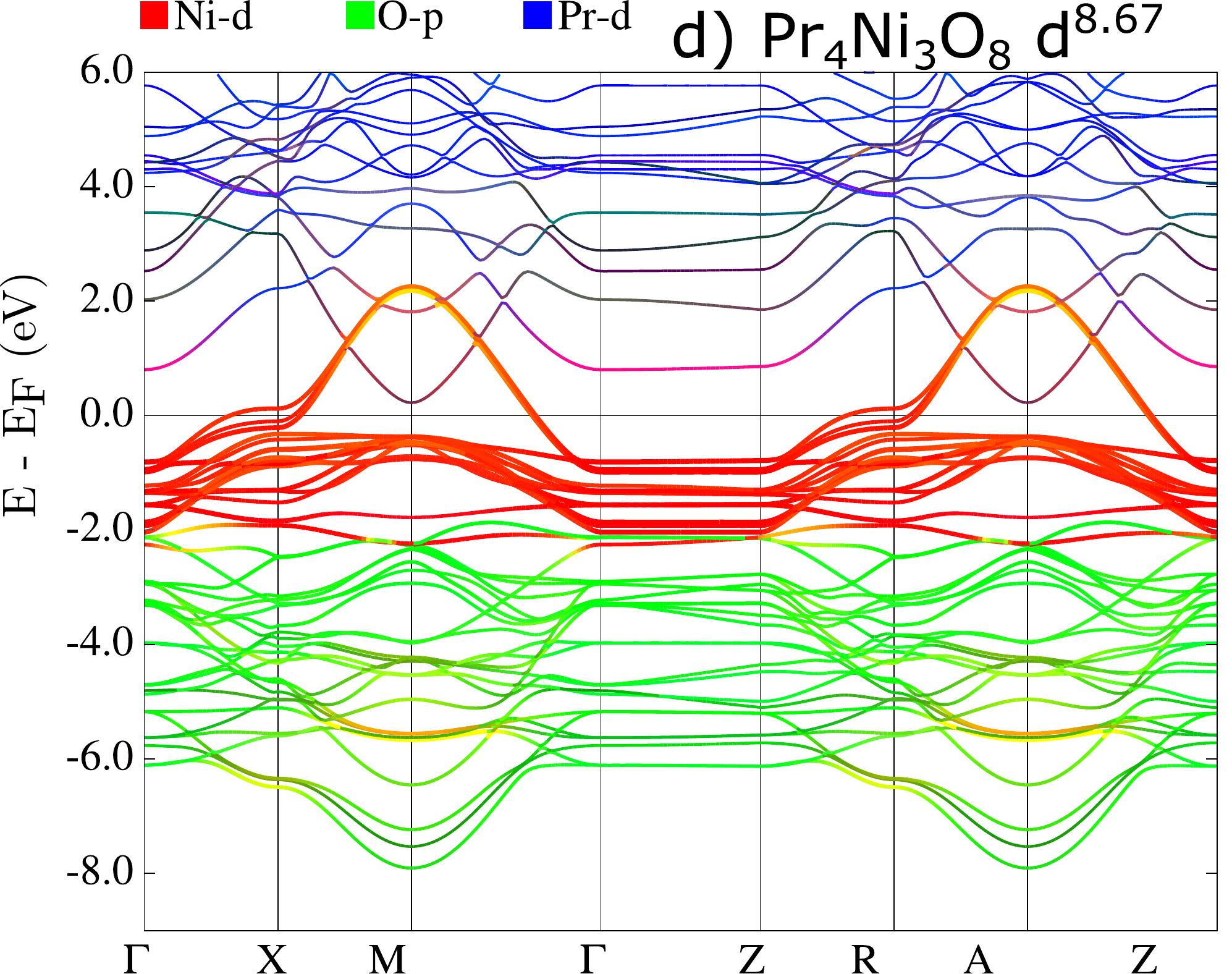}}
    \caption{DFT band structures plotted along high symmetry lines of the primitive tetragonal unit cell with orbital character shown in color for \NNO{} $d^9$ (top left), \NNO{} $d^{8.67}$ (top right), \PNO{} $d^9$ (bottom left), and \PNO{} $d^{8.67}$ (bottom right).}
    \label{fig:dft_bands}
\end{figure*}

\begin{figure}[t]
    \centering
    \includegraphics[width = \linewidth]{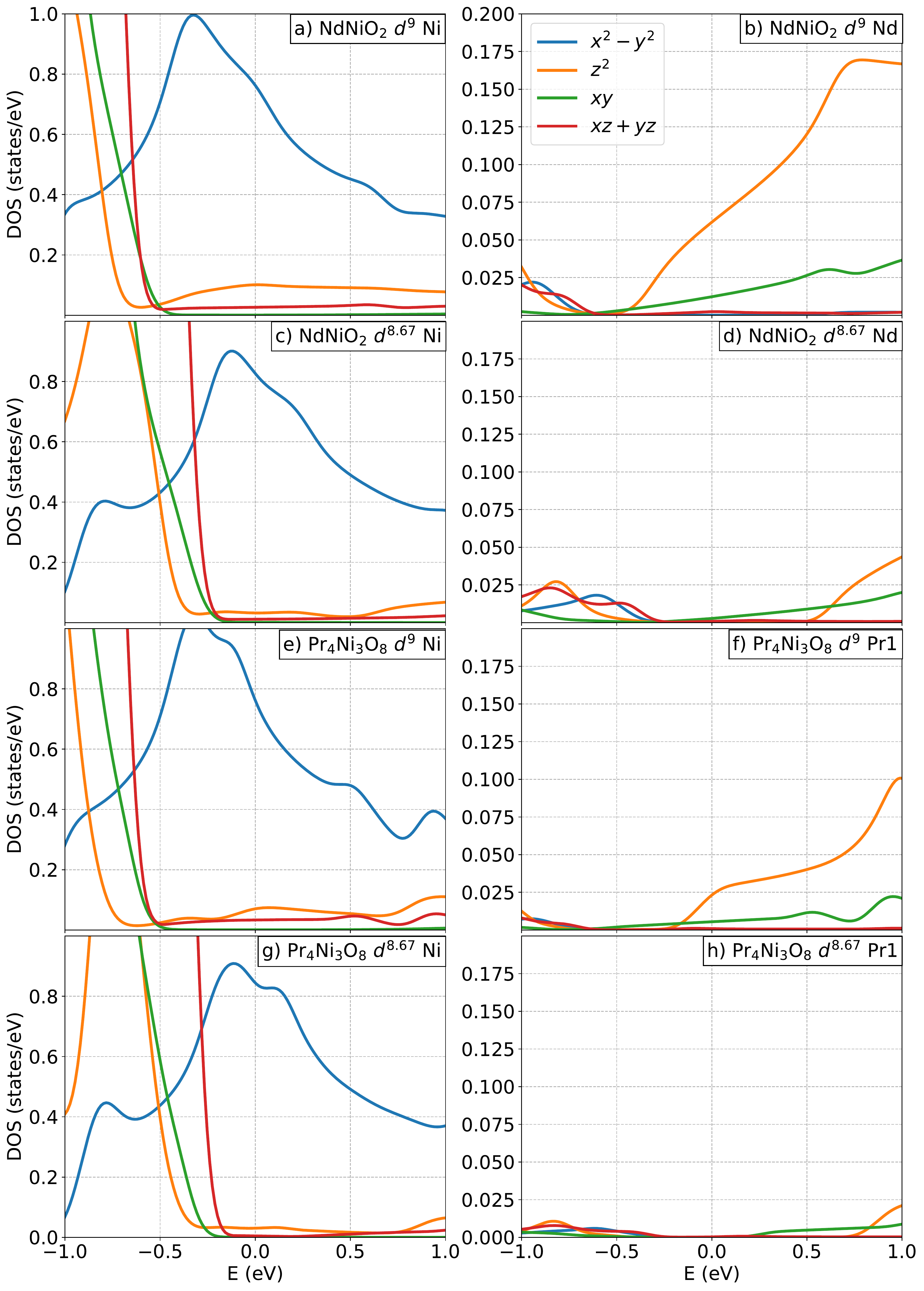}
    \caption{Near Fermi energy DFT density of states per Ni atom resolved into Ni-$d$ and Nd/Pr1-$d$ orbitals.}
    \label{fig:dft_dos}
\end{figure}

Fig.~\ref{fig:dft_bands} shows the DFT band structures along high symmetry directions for both materials at the two nominal fillings in the non-magnetic state. Using the Prima package \cite{primapy}, we show the Ni-$d$, O-$p$ and Nd/Pr1-$d$ character of the bands. In Appendix~\ref{app:fat_bands}, we show the specific orbital character of the near Fermi energy bands. First, we point out that the overall band structures are very similar when compared at the same doping level. For both materials, the main active bands crossing the Fermi level are of Ni-$d_{x^2-y^2}$ character with some admixture of O-$p$ (see also Fig.~\ref{fig:dft_dos}). The other Ni-$d$-derived bands are lower in energy, and the O-$p$-derived bands even lower, implying a relatively large charge-transfer energy and, as discussed in previous work \cite{Lee04, botana2019, nomura2019, lechermann2019late, choi2019role, jiang2020critical}, placing the materials closer to the Mott-Hubbard than to the charge transfer regime. 

The Nd/Pr-$d$-derived bands are mostly above the Fermi energy and weakly hybridized with the Ni-$d$  bands, specifically d$_{z^2}$ near $\Gamma$ and d$_{xz/yz}$ near A (and M in the case of the trilayer compound), as extensively discussed in previous works \cite{Lee04, karp2020manybody, botana2019, nomura2019, hepting2019, wu2019, sakakibara2020model, gao2019electronic, zhang2019effective, jiang2019electronic, hirayama2019materials, Gu2020substantial, si2020topotactic, choi2019role, Liu2020electronic, olevano2020abinitio, leonov2020lifshitz, Kitatani2020nickelate}. Depending on the doping, these states can give rise to small Fermi surface sheets centered at $\Gamma$ and $A(/M)$. In the case of \PNO{}, the near Fermi energy contributions to the electronic structure come from Pr1 and not Pr2. An overall difference between the two materials is that the \PNO{} bands do not have much $k_z$ dispersion because of  the fluorite blocks and body centered tetragonal shift mentioned above. 
 
 Comparison of the left and right columns of Fig.~\ref{fig:dft_bands} shows that as the nominal Ni valence is reduced from $d^9$ to $d^{8.67}$,  the charge transfer energy is reduced:  the O-$p$-derived bands move closer to and entangle more with the Ni-$d$-derived bands. We quantify the charge transfer energies by projecting the DFT bands onto maximally localized Wannier functions \cite{MLWF1, MLWF2} using Wannier90 \cite{wannier90_v3,wien2wannier}. We then take the charge transfer energy as the difference between the on-site energy for the Ni-$d_{x^2-y^2}$ and the O-$p_{\sigma}$ orbital within the same NiO$_2$ plane. The resulting charge transfer energies are shown in Table \ref{tab:charge_transfer} and vary by less than $5\%$ between the two materials at the same doping, and by less than $10\%$ over the doping range considered. Comparison to previous DFT calculations \cite{nica2020theoretical} indicates that the absolute values of the charge transfer energies are somewhat dependent ($\sim \SI{0.2}{eV}$) on rare earth ion, with larger $Z$ ions having a larger charge transfer energy.

Looking now in more detail at the Nd/Pr1-derived bands, we see that these bands are weakly hybridized with the Ni-$d$ bands, the Nd/Pr1-$d_{z^2}$ mainly with Ni-$d_{z^2}$ and the Nd/Pr1-$d_{xy}$ mainly with Ni-$d_{xz/yz}$. In both our DFT and DFT+DMFT calculations, both bands cross the Fermi energy in the $d^9$ case, but as carriers are removed the bands empty out. For $d^{8.67}$, the $\Gamma$ centered band is above the Fermi energy in both materials; for the $d^{8.67}$ infinite layer material, a small $A$ centered pocket remains, but in the trilayer material the $A/M$-centered pocket disappears. The difference arises from the difference in c-axis dispersions, and as discussed below may be relevant to the low energy physics (i.e., charge/spin order). 

\begin{table}[]
\begin{tabular}{|l|l|l|}
\hline
                              & $d^9$ & $d^{8.67}$ \\ \hline
\NNO{}       & 4.35  & 3.81       \\ \hline
\PNO{} inner & 4.32  & 4.01       \\ \hline
\PNO{} outer & 4.32  & 3.89       \\ \hline
\end{tabular}
\caption{Charge transfer energies (in eV) obtained from Wannier fits to the DFT band structure. The Ni-$d$, O-$p$, and Nd/Pr1-$d_{z^2}$ and $d_{xy}$ orbitals are included in the fit. The charge transfer energy is defined as the difference between the onsite energies of the Ni-$d_{x^2-y^2}$ and O-$p_\sigma$ Wannier functions.}
\label{tab:charge_transfer}
\end{table}

\section{DMFT Results \label{sec:DMFTSelf}}

\subsection{Self energies and mass enhancements}
\begin{figure}[t]
    \centering
    \includegraphics[width = \linewidth]{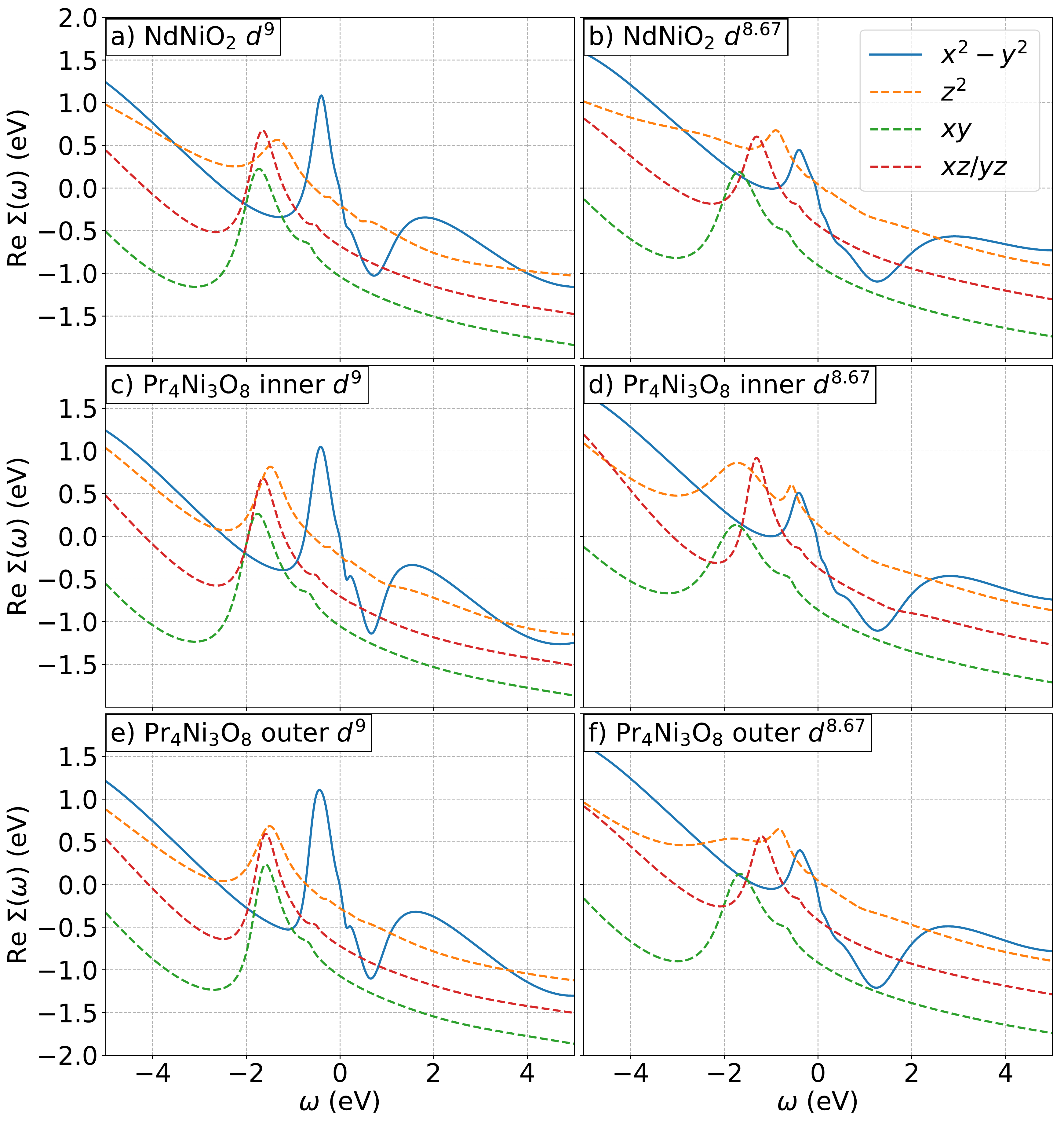}
    \caption{Real part of the analytically continued self energies of the correlated Ni-$d$ orbitals for undoped \NNO{} (top left), 1/3 hole doped \NNO{} (top right), 1 electron doped \PNO{} (middle and bottom left), and undoped \PNO{} (middle and bottom right).  The double counting and chemical potential are subtracted from the self energies.}
    \label{fig:sigma}
\end{figure}

Fig.~\ref{fig:sigma} shows  the real part of the analytically continued self energies. The self energies of the two materials are very similar when compared at the same doping. The $d_{x^2-y^2}$ self energy has substantial structure in the $-\SI{1}{eV}\lesssim \omega \lesssim \SI{1}{eV}$ near Fermi energy range; the other orbitals have a much smoother self energy in this range, confirming that the $d_{x^2-y^2}$ orbital is the dominant correlated orbital in these materials.

We quantify the strength of electronic correlations by the inverse quasiparticle renormalization $Z^{-1}=1-\partial Re\Sigma(\omega\rightarrow 0)/\partial \omega$ related, in the single-site DMFT approximation, to the quasiparticle mass enhancement as $m^\star/m=Z^{-1}$.  Results are shown in Table \ref{tab:mass}; they confirm further that the correlations are dominated by the $d_{x^2-y^2}$ orbitals. We also see that the two materials have very similar $d_{x^2-y^2}$ mass enhancements at the same doping and that the effective correlation strength is somewhat greater at $d^9$ filling. 

The $d_{x^2-y^2}$ self energy has pronounced structures at $\sim \pm \SI{0.5}{eV}$; these structures are a consequence of Mott-Hubbard/charge transfer correlations, and at the Mott transition would coalesce near $\omega=0$. The structures are much more pronounced at $d^9$, indicating the weakening of correlations upon hole doping expected of Mott-Hubbard/charge transfer materials. We also note the presence of a small structure at about $\omega= \SI{0.2}{eV}$ visible especially in the $d^9$ case; this might be a signature of Hund's physics because a low frequency structure observed on only one side of the Fermi energy is characteristic of known Hund's metal materials \cite{karp2020SMO}.   Indeed, some authors label \NNO{} as a Hund's metal \cite{wang2020hunds, kang2020infinitelayer, kang2020optical}. The peak we find, along with the $d$-level density matrix discussed below, may be an indication that Hund's metal physics plays at least some role; however, the small amplitude of the feature and its presence only in the $d_{x^2-y^2}$ self energy and not in the self energy of other orbitals, and visible only at $d^9$, suggests that the Hund's correlations, while present, are less important than the Mott-Hubbard correlations revealed by the large amplitude features in the self energy. The relative strength of Hund's versus Mott-Hubbard physics depends on the ratio of $U$ to $J$, and comparing to results presented in Ref.~\cite{ryee2020unveiling}, especially the value of $Z$, our results are more on the Mott-Hubbard side of the $U$-$T$ phase diagram. 

\begin{table}[]
\begin{tabular}{|l|l|l|l|l|}
\hline
                                       & $d_{z^2}$ & $d_{x^2-y^2}$ & $d_{xy}$ & $d_{xz/yz}$ \\ \hline
NdNiO$_2$ $d^9$                      & 1.4           & 3.7       & 1.4      & 1.3         \\ \hline
NdNiO$_2$ $d^{8.67}$                   & 1.4           & 2.9       & 1.4      & 1.4         \\ \hline
Pr$_4$Ni$_3$O$_8$ $d^{8.67}$ inner        & 1.5           & 3.0       & 1.4      & 1.4         \\ \hline
Pr$_4$Ni$_3$O$_8$ $d^{8.67}$ outer        & 1.4           & 3.0       & 1.4      & 1.4         \\ \hline
Pr$_4$Ni$_3$O$_8$ $d^9$ inner & 1.4           & 3.9       & 1.4      & 1.4         \\ \hline
Pr$_4$Ni$_3$O$_8$ $d^9$ outer & 1.4           & 4.0       & 1.4      & 1.3         \\ \hline
\end{tabular}
\caption{Mass Enhancements for \PNO{} and \NNO{} at studied doping levels, resolved by orbital character. The values are extracted from the Matsubara self energies, as described in Appendix~\ref{app:dmft}.}
\label{tab:mass}
\end{table}

\subsection{Spectral Functions}

\begin{figure}[t]
    \centering
    \includegraphics[width = \linewidth]{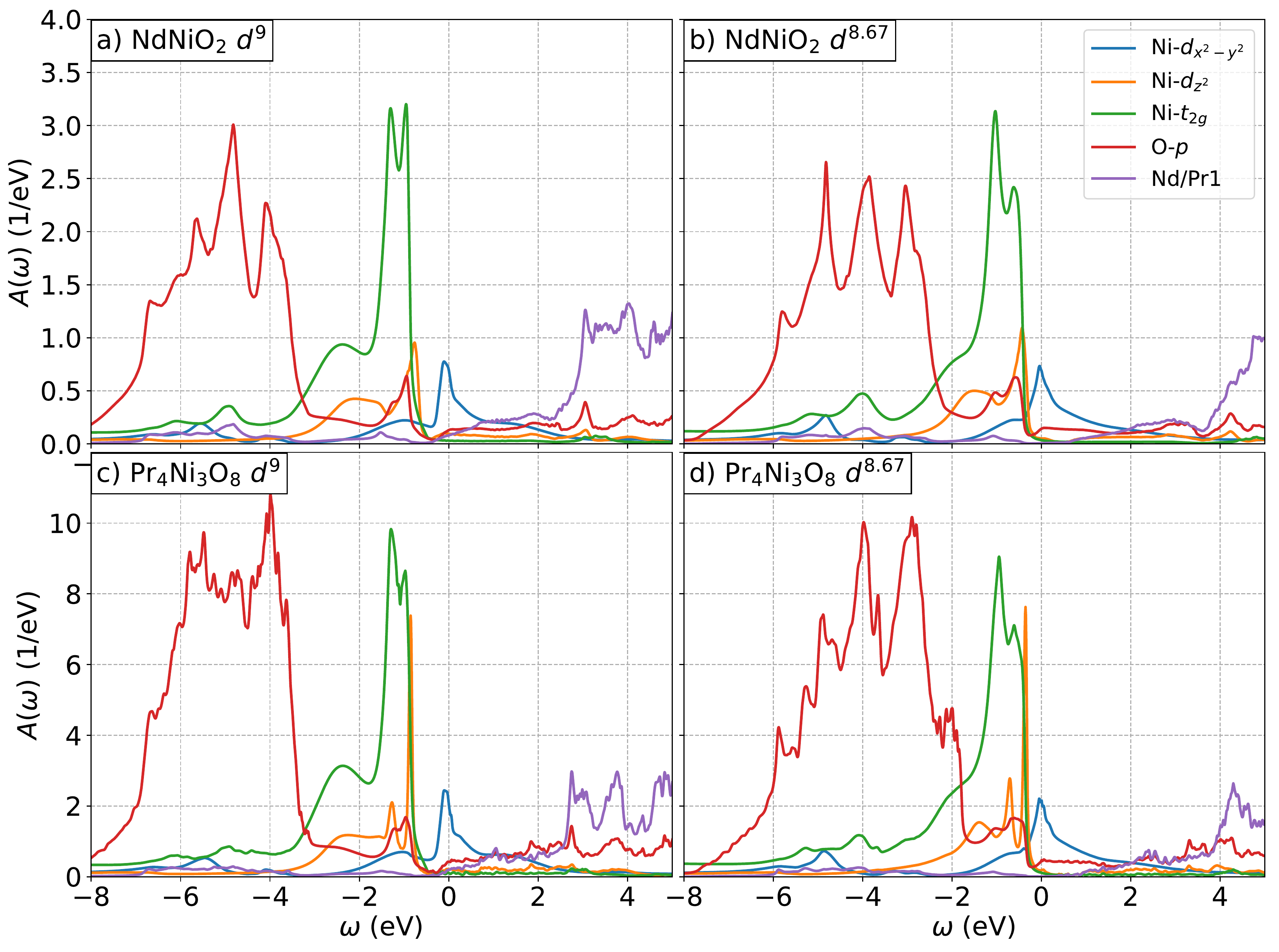}
    \caption{DFT+DMFT spectral functions (summed over spin) for undoped \NNO{} (top left), 1/3 hole doped \NNO{} (top right), 1 electron doped \PNO{} (bottom left), and undoped \PNO{} (bottom right).}
    \label{fig:Aw}
\end{figure}

Fig.~\ref{fig:Aw} shows the orbitally resolved DFT+DMFT spectral function ${\bf A}(\omega) = i\left[{\bf G}(\omega) - {\bf G}(\omega)^\dag\right]/2\pi$ (${\bf A}$ and ${\bf G}$ are matrices in orbital space). The spectral functions for the two materials are similar when compared at the same nominal carrier concentration. One difference between the two materials is that in the $d^{8.67}$ case for \PNO{}, there is a weak shoulder in the O-$p$ spectral function at $\sim \SI{-2}{eV}$ that is not present for $d^{8.67}$ \NNO{}. This is due to the oxygen atoms in the fluorite block of \PNO{} which are not present in \NNO{}. The charge transfer energies discussed above are visualized qualitatively here as the energy separation between the $d$ and $p$ densities of states. It can be observed that, consistent with the previous discussion, upon hole doping the oxygen states move somewhat closer to the Ni-$d$ states. 

\begin{figure}[t]
    \centering
    \includegraphics[width = \linewidth]{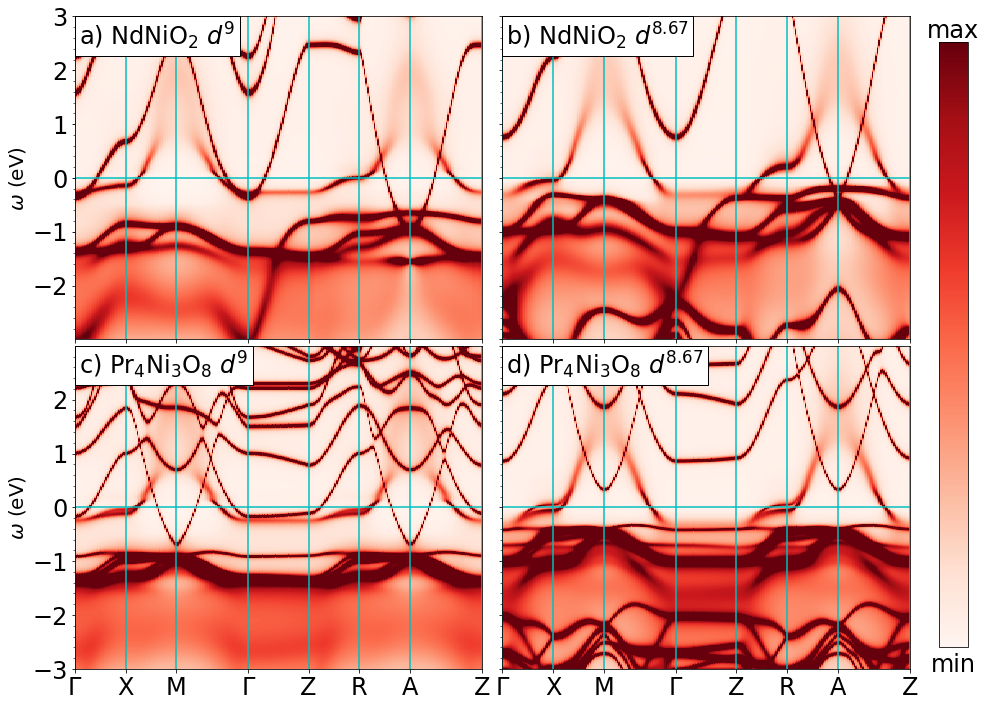}
    \caption{DFT+DMFT momentum-resolved spectral functions $A(k,\omega)$  per Ni atom for undoped \NNO{} (top left), 1/3 hole doped \NNO{} (top right), 1 electron doped \PNO{} (bottom left), and undoped \PNO{} (bottom right).}
    \label{fig:Akw}
\end{figure}

Fig.~\ref{fig:Akw} shows the momentum-resolved spectral functions $A(k,\omega)= -Tr\left[\text{Im}\mathbf{G}(k,\omega)\right]/\pi$ along the same high-symmetry paths in the Brillouin zone used  to plot the DFT band structure (Fig.~\ref{fig:dft_bands}). The many-body electronic structure is well represented as a set of bands, renormalized from the DFT values by correlations. Comparison of the bands for the two materials reveals broad similarities, but some important differences in detail. For the trilayer material, the absence of c-axis hopping between the three-layer structural units means that the $k_z$-dispersing bands visible in the infinite layer case appear as a triple of $k_z$-independent bands for the trilayer material (compare the $\Gamma\rightarrow Z$ dispersions of the two material families). As in the DFT case, at the $d^9$ valence, the Nd/Pr1 $5d$ band at $A$ is somewhat less deep for the trilayer material than for the infinite-layer one, and its Fermi surface is eliminated completely for the trilayer but not the infinite-layer case at $d^{8.67}$. Finally,  the $k_z$-dispersion for \NNO{} means that the van Hove singularities at $(\pi,0)$ crosses the Fermi energy only at one $k_z$ value, whereas for \PNO{} the $k_z$ dispersion is negligible but there are three discrete van Hove singularities (two below, one above $E_F$). These differences in Fermiology may be relevant for low energy instabilities, as we mention below.

 \begin{figure*}
    \centering
    \includegraphics[width = \linewidth]{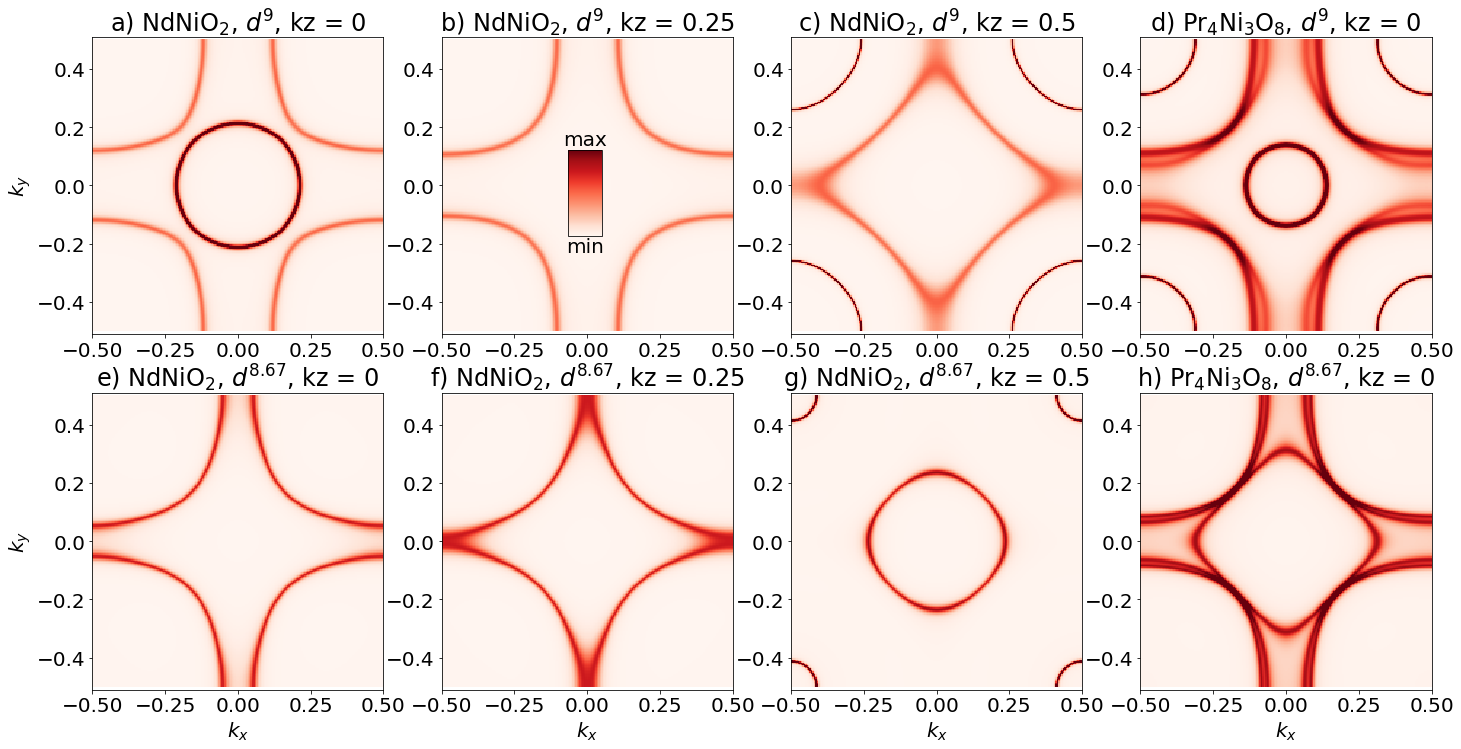}
    \caption{DFT+DMFT momentum-resolved spectral functions $A(k_{x,y},k_z,\omega=0)$ for $d^9$ (top row) and $d^{8.67}$ (bottom row). Three left panels: \NNO{} at different $k_z$ values indicated; right panel: \PNO{}.  $k$ is given in 2$\pi$/a and 2$\pi$/c units.}
    \label{fig:mbfs}
\end{figure*}

Fig.~\ref{fig:mbfs} shows the many-body Fermi surface, defined as the many-body spectral function evaluated at $\omega=0$ and, for \NNO{}, at several $k_z$.  Panels (a-c) show that as $k_z$ is increased from $0$ in the stoichiometric infinite layer material, the $\Gamma$-centered Nd-derived pocket vanishes and is replaced by an A-centered pocket, while the Fermi surface of the  Ni-$d_{x^2-y^2}$-derived band passes through a van Hove singularity, changing its topology from hole-like ($M$-centered) to electron-like ($\Gamma$-centered). The \PNO{} Fermi surface has negligible $k_z$ dispersion, but as shown in panel (d) at nominal $d^9$ carrier concentration consists of three Ni-$d_{x^2-y^2}$-derived pockets: the bonding, non-bonding, and antibonding superpositions of the three layers, which all have the same hole-like topology. Pr-derived pockets at $\Gamma$ and $A/M$ are also evident. 

For \NNO{} at nominal $d^{8.67}$ doping, we see from panels (e-g) that the van Hove singularity in the Ni-derived bands occurs at a smaller $k_z$ and near $k_z=\pi$ the Fermi surface cross section becomes a small $\Gamma$-centered circle. The $\Gamma$-centered Nd-derived pocket is absent and the $A$-centered Nd-derived pocket is much smaller. Turning now to the trilayer material (panel h), we see that at nominal $d^{8.67}$ filling both Pr-derived bands are above the Fermi energy, while the $d_{x^2-y^2}$ antibonding Fermi around $\Gamma$ surface becomes nearly square-like with a nesting vector similar (but not equal) to the observed density wave ordering vector \cite{Zhang2016stacked}.  DFT studies have indeed revealed that the free energy is lowered if a density wave at $q=(2\pi/3,2\pi/3)$ is considered \cite{botana2016charge, Zhang2017}.

\subsection{Orbital Occupancies and Occurrence Probabilities}

\begin{table}[]
\begin{tabular}{|l|l|l|l|l|l|}
\hline
                                       & $d_{z^2}$ & $d_{x^2-y^2}$ & $d_{xy}$ & $d_{xz/yz}$ & total \\ \hline
NdNiO$_2$ $d^9$                      & 1.59      & 1.13          & 1.96     & 1.91        & 8.51  \\ \hline
NdNiO$_2$ $d^{8.67}$                   & 1.64      & 1.03          & 1.98     & 1.93        & 8.50  \\ \hline
Pr$_4$Ni$_3$O$_8$ $d^{8.67}$ inner        & 1.63      & 1.03          & 1.97     & 1.93        & 8.49  \\ \hline
Pr$_4$Ni$_3$O$_8$ $d^{8.67}$ outer        & 1.66      & 1.02          & 1.97     & 1.93        & 8.53  \\ \hline
Pr$_4$Ni$_3$O$_8$ $d^9$ inner & 1.59      & 1.14          & 1.96     & 1.91        & 8.52  \\ \hline
Pr$_4$Ni$_3$O$_8$ $d^9$ outer & 1.62      & 1.14          & 1.96     & 1.91        & 8.54  \\ \hline
\end{tabular}
\caption{Ni-$d$ orbital occupancies obtained from the impurity ${\bf G}(i\omega_n)$.}
\label{tab:fillings}
\end{table}

Table \ref{tab:fillings} shows the fillings of the correlated orbitals, obtained from the impurity Green's function ${\bf G}(i\omega_n)$, for both materials at both dopings. Remarkably, for both materials, within the context of our 5 orbital model, removing 1/3 of an electron/Ni leaves the total occupancy of the Ni-$d$-states roughly invariant. The $d_{x^2-y^2}$ occupation decreases as expected, but this is mostly compensated by an increase in the occupation of the other orbitals.  In this context, it should be remembered that there is some admixture of O-$p$ and Pr/Nd-$d$ in our effective orbital basis used in DMFT \cite{wang2020hunds}. 

\begin{table}[]
\begin{tabular}{|l|l|l|l|l|l|}
\hline
                                   & $d^7$ & LS $d^8$ & HS $d^8$ & $d^9$ & $d^{10}$ \\ \hline
NdNiO$_2$ $d^9$                    & 0.05  & 0.13     & 0.29     & 0.49  & 0.04     \\ \hline
NdNiO$_2$ $d^{8.67}$               & 0.05  & 0.17     & 0.25     & 0.48  & 0.04     \\ \hline
Pr$_4$Ni$_3$O$_8$ $d^{8.67}$inner  & 0.05  & 0.17     & 0.26     & 0.48  & 0.04     \\ \hline
Pr$_4$Ni$_3$O$_8$ $d^{8.67}$ outer & 0.05  & 0.17     & 0.24     & 0.50  & 0.04     \\ \hline
Pr$_4$Ni$_3$O$_8$ $d^9$ inner      & 0.05  & 0.12     & 0.29     & 0.49  & 0.04     \\ \hline
Pr$_4$Ni$_3$O$_8$ $d^9$ outer      & 0.05  & 0.12     & 0.28     & 0.51  & 0.05     \\ \hline
\end{tabular}
\caption{Occurrence probabilities of different Ni $d$ valence states obtained from the impurity density matrix found in the fully charge self consistent calculations. For $d^8$ we further decompose the probabilities into low spin (LS; $S=0$) and high spin (HS, $S(S+1)=2$). Other valence configurations occur with negligible probabilities.}
\label{tab:occ_prob}
\end{table}

Table \ref{tab:occ_prob} shows the occurrence probabilities of different configurations of the Ni-$d$ states obtained from the impurity density matrices determined from the CTHYB solver. We find similar results for both materials, and these results are only weakly dependent on doping. We find that the materials have $\approx 50\%\ d^9$, and the rest is mostly $\approx 40\%$ $d^8$ with  $\approx 6\%$ $d^7$ and $\approx 4\%$ $d^{10}$. Approximately $70\%$ of the $d^8$ weight is high spin for the nominal $d^9$ filling calculations, decreasing to $\approx 60\%$ for nominal $d^{8.67}$ filling. In all cases, the contribution to  high  spin $d^8$ is mainly from one electron in $d_{x^2-y^2}$  and the other in $d_{z^2}$. If only the $d_{x^2-y^2}$ orbital were relevant, as in traditional one band Mott-Hubbard systems, we would expect equal amounts of $d^{10}$ and $d^8$ for the nominal $d^9$ materials. If there were more $d^{10}$ than $d^8$, then like cuprates \cite{Eskes88, karp2020manybody} there would be charge transfer from the oxygen orbitals. In this case, in stark contrast to cuprates, we find much more $d^8$ than $d^{10}$, indicating a reverse charge transfer from Ni to Nd/Pr. We should also remark that the stripe state seen for the La variant of the trilayer material is consistent with non-magnetic domain walls, implying they are occupied by low-spin $d^8$ \cite{Zhang2017, zhang2019spin, lin2020strong} as supported by DFT studies \cite{botana2016charge,Zhang2017}.

Interestingly, even though the total Ni-$d$ occupancy is essentially independent of doping and the $d_{x^2-y^2}$ occupancy gets closer to half filling as electrons are removed, the $d_{x^2-y^2}$ self energy evolves with doping roughly as expected in a doped Mott-Hubbard material being largest at nominal $d^9$ and decreasing as carriers are removed. This finding is consistent with other 5 orbital DFT+DMFT studies \cite{wang2020hunds} and suggests that the effective low energy Mott-Hubbard physics arises in an interesting way from charge transfer physics. In contrast to the cuprates, where the ligand (oxygen) states both provide bandwidth for the $d$ orbitals and act as a charge reservoir, absorbing most of the doping, in the nickelate materials the charge reservoir is provided by the Nd/Pr $5d$ states but these orbitals do not provide the $d$ bandwidths. Instead, the bandwidth is mainly due to hybridization with O-$p$ as in cuprates, which is supported by resonant x-ray inelastic (RIXS) studies showing a strong fluorescence line due to $d$-$p$ mixing \cite{hepting2019}.

The occurrence probabilities are different from those obtained in our previous work on \NNO{} \cite{karp2020manybody}, where we found 0.05 probability of $d^8$ and 0.26 probability of $d^{10}$. The first source of difference comes from the differently constructed low energy subspaces: by considering $5$ rather than $2$ $d$-orbitals we provide more possibilities for $d^8$ configurations.
The second difference is methodological, and points to an interesting issue in the DFT+DMFT formalism. In this work we use a projector formalism, while in the previous work we used the selectively localized variant of the maximally localized Wannier function method. The bands obtained from the selectively localized Wannier procedure reproduce the DFT bands perfectly. However, the physical content of the  orbital basis in which the correlated problem is solved differs between methods. The oxygen and Pr/Nd Wannier functions defined in the selectively localized procedure overlap in space with the Ni-$d$ orbitals; some of this ``ligand" amplitude appears as Ni amplitude in the projector methodology. This difference in how the methods disentangle the Ni, Nd, and O contributions leads to the differences in $d$ occupancy; it is important to note that the differences are larger for the $d$ occupancies than for other quantities: the two methods give very similar mass enhancements and lifetimes for the near Fermi energy Ni-$d_{x^2-y^2}$-derived bands. The choice of correlated orbitals is a fundamental ambiguity in the DFT+DMFT methodology that requires further investigation \cite{aichhorn2009}.

\section{Discussion}\label{sec:Conclusion}

In this paper, we presented DFT+DMFT studies of \NNO{} and \PNO{}, representative of two families of cuprate-analog materials involving square planar coordinated near $d^9$ valence Ni ions but with other structural differences that lead to different Fermiology. In terms of formal valence, \PNO{} corresponds to 1/3 hole doped \NNO{}.  The \PNO{} family of materials has not previously been studied with DFT+DMFT. 

Our study employed the DFT+DMFT method, with full charge self-consistency to properly account for the charge transfer between the Ni and the Nd/Pr ions; this is believed to correctly incorporate the physics of the long-ranged Coulomb interaction in moderating density inhomogenieties in solids~\cite{hampel:2019, Chen:2015}.

Our study was based on treating correlation effects in a wide energy window; instabilities arising from very low energy physics is beyond the scope of our study. 

In our study, we found that the materials have very similar electronic properties on the broader energy scales when compared at the same doping level: from the point of view of basic strong correlation many-body physics, the two compounds may be studied interchangeably. Our calculations indicate that differences of physics should be attributed to low energy physics arising from differences in Fermiology.

On the level of broad-band electronic structure we found, in agreement with previous work \cite{hepting2019, kang2020infinitelayer,werner2019nickelate,lechermann2019late, wang2020hunds}, that the ground state electronic configuration has a significant admixture of $d^8$ and a relatively small admixture of $d^{10}$, unlike the cuprates where the ground state electronic configuration is an almost equal admixture of $d^9$ and $d^{10}\underbar{L}$ (\underbar{L} denotes a ligand hole) with very little $d^8$. The appearance of $d^8$ without $d^{10}$ is a consequence of charge transfer from the Ni to the Nd/Pr orbitals,  which as other authors have noted function as a charge reservoir. However, we  found that the self energy and the spectral function displayed the characteristic forms expected in a Mott-Hubbard/charge transfer system, including (for the more strongly correlated nominal d$^9$ valence)  the characteristic three-peak structure in the spectral function  and a self energy characterized by roughly particle-hole symmetric structures $\pm \SI{0.5}{eV}$ above and below the Fermi energy.  These features, along with the modest ($\lesssim 30\%$) admixture of high-spin $d^8$ into the ground state, imply that Hund's metal physics arising from the high-spin $d^8$, while potentially present, does not play a large role in the basic correlation physics. Rather, correlations are dominated by the $d_{x^2-y^2}$ orbital which makes by far the most important contribution to the near $E_F$ density of states and is much more strongly correlated than the other $d$ orbitals, suggesting that a one band plus charge reservoir Mott-Hubbard-like description of the low energy physics may be more appropriate.  We emphasize, though, that a proper inclusion of the Nd/Ni $d_{z^2}$ charge reservoir bands is essential.

These results have implications for the downfolding of our model to a few band low energy model.  The relatively low weight of oxygen states along with the presence of Nd-derived bands at the Fermi level means that, in contrast to the cuprate where there is a still not fully settled debate on the dynamical significance of oxygen states for the low energy physics, for the nickelate materials the question is whether the physics is of a one band Hubbard model plus a charge reservoir or whether a two band (Ni-$d_{x^2-y^2}$ and Ni-$d_{z^2}$) Hubbard model plus reservoir is a more correct description. The results presented here favor the first interpretation but further investigation of this question is of interest.

We found differences in the low energy physics, related in particular to the presence or absence of a rare-earth derived pocket near the $A$ point of the Brillouin zone, and to the specifics of the van Hove singularities associated with the Ni-$d_{x^2-y^2}$-derived bands.  In particular the antibonding Fermi surface for the trilayer material exhibits a nesting vector near to the observed density wave vector. The nesting will lead to a peak in the susceptibility at the nesting vector, which will favor density wave ordering at or near this wavevector. On the other hand, the observed density wave state resembles that seen in La$_{2-x}$Sr$_x$NiO$_4$ near $x$=1/3 which is due to real space diagonal stripes \cite{Zhang2016stacked};  further studies of the density wave ordering  are in progress. Regardless,  our general finding that the significant electronic structure differences between the trilayer and infinite layer nickelates relate to the Fermiology means that future experimental and theoretical studies should give us a better picture of the relation of Fermi surface-driven and local physics in this class of materials.  

To conclude, we hope that our work will set the stage for a detailed comparison of experimental results on the two material families.  Our results demonstrate the importance of making comparisons at the same doping level. For example, a recent resonant x-ray scattering study \cite{lin2020strong} found, from the spin-wave dispersions, that the trilayer nickelate materials have a superexchange strength that is about three times larger than that estimated in the infinite layer material from Raman scattering \cite{fu2019corelevel}. Treatment of the superexchange is beyond the scope of the present paper, though it should be suppressed as the charge transfer energy increases when going from $d^{8.67}$ to $d^9$. Further motivation for performing experiments at a similar carrier concentration comes from the observation that, as of yet, superconductivity has not been reported in the trilayer nicklate family, probably because the doping of the stoichiometric materials is too high, while magnetism, in the form of a density wave instability, has been observed in the trilayer but not the infinite-layer nickelates.  Investigations of the possibility of superconductivity in electron doped \PNO{} and a density wave instability in hole doped \NNO{} should provide further insight on the interplay of Fermiology and local correlation physics.

\section{Acknowledgements}
J.K., A.J.M., M.R.N. and H.P. acknowledge funding from the Materials Sciences and Engineering Division, Basic Energy Sciences, Office of Science, US DOE. AB acknowledges NSF-DMR grant 1904716. The Flatiron Institute is a division of the Simons Foundation.

\appendix

\section{DFT+DMFT Calculations}
\label{app:dmft}

\begin{figure*}[t]
    \centering
    \includegraphics[width = \linewidth]{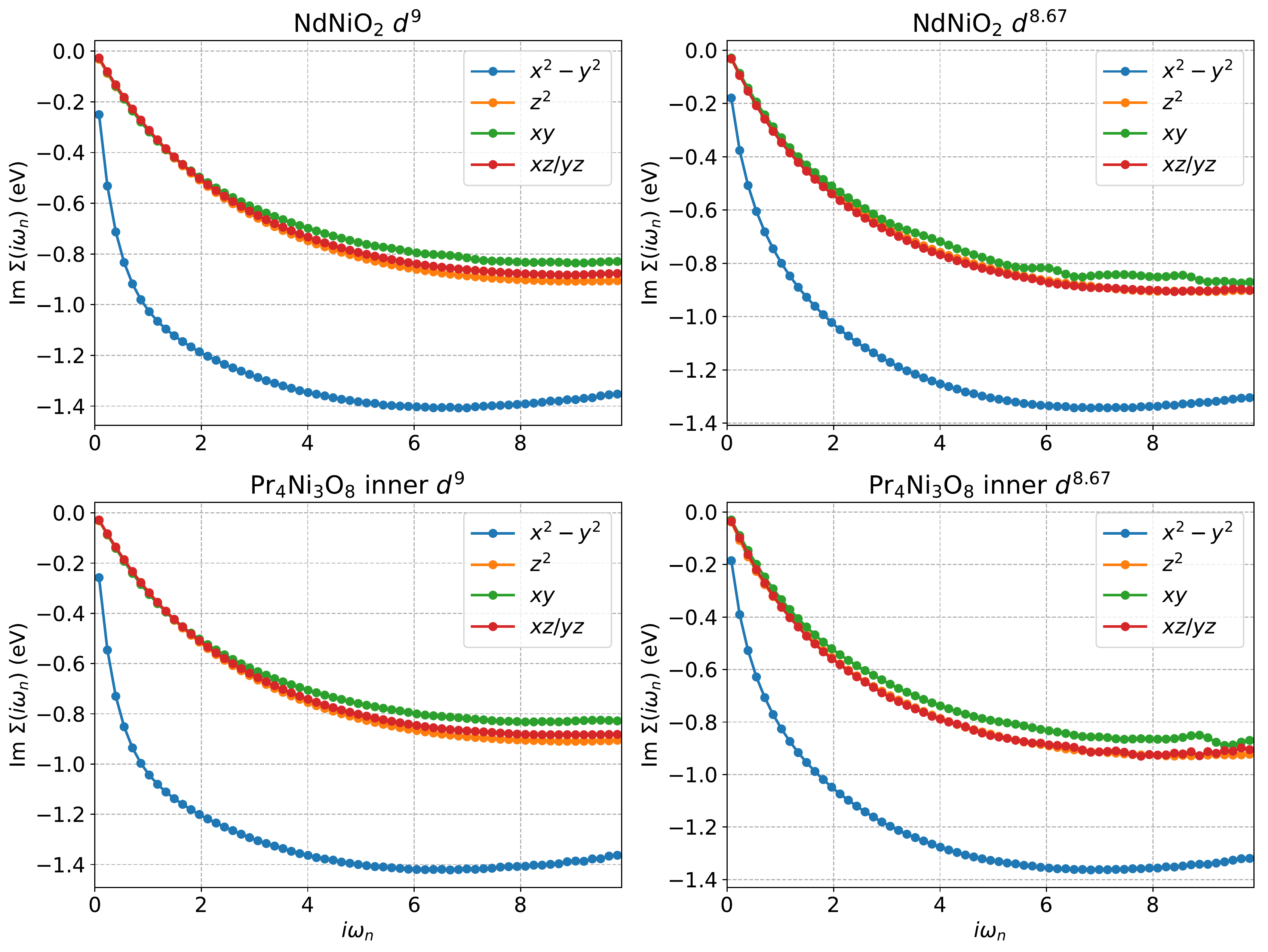}
    \caption{Imaginary part of the Matsubara self energies.}
    \label{fig:Im_Sigma_iw}
\end{figure*}

We perform DFT calculations using WIEN2k~\cite{Blaha2018} with the standard PBE version of the GGA functional~\cite{PBE}. For both materials we use the experimentally determined crystal structure. For \NNO{} this is the P4/mmm space group with $a = b = \SI{3.92}{\angstrom}$ and $c = \SI{3.31}{\angstrom}$ \cite{Hwang2019}. For \PNO{} this is the I4/mmm space group with $a = b = \SI{3.9347}{\angstrom}$ and $c = \SI{25.485}{\angstrom}$ \cite{Zhang2017}. The DFT calculations are converged with a $RK_{max}=7$ and with a $k$-point grid of $40 \times 40 \times 40$ for \NNO{} and $20 \times 20 \times 20$ for \PNO{}. We put the Nd/Pr-$4f$ bands in the core. We dope the materials using the virtual crystal approximation, where we adjust the atomic numbers of the Nd/Pr ions to fractional values and correspondingly change the number of electrons. For the DMFT calculations we construct projectors in an energy window of $\SI{-10}{eV}$ to $\SI{10}{eV}$ around the Fermi energy. 

We perform the DMFT calculations using the TRIQS software library \cite{TRIQS, TRIQS/CTHYB, TRIQS/DFTTOOLS}. We treat the 5 Ni-$d$ orbitals as correlated. We use a rotationally invariant Slater Hamiltonian with $U = F^0 = \SI{7}{eV}$ and $J = (F^2 + F^4)/14 = \SI{0.7}{eV}$. We perform the calculations at a temperature of $T = \SI{290}{K}$. We solve the impurity problem using the CTHYB solver \cite{gull2011ctqmc, TRIQS/CTHYB}. We use a double counting correction of the FLL form \cite{liechtenstein1995fll}, which we update at each iteration as the DFT density changes. We analytically continue the self energies using the maximum entropy method \cite{TRIQS/maxent}. 

\subsection{Matsubara Self Energy and Mass Enhancement}

Fig.~\ref{fig:Im_Sigma_iw} shows the imaginary part of the Matsubara self energies for both materials at their nominal fillings. The Matsubara self energies clearly show that correlations are stronger for the $d_{x^2-y^2}$ orbital and similar in strength for the other orbitals. 

We obtain the quasiparticle mass enhancement directly from the Matsubara self energy to avoid error in the analytic continuation. The mass enhancement is given by:
\begin{equation}
Z^{-1} = \left( 1 - \frac{\partial \text{Im}\Sigma(\imag \omega_n)}{\partial \omega_n}\Big|_{\omega_n \to 0}\right).
\end{equation}
We determine $Z$ by fitting a polynomial of fourth order to the lowest six points of the Matsubara self-energies and extrapolate Im$\left[\Sigma(\imag\omega_n\rightarrow 0)\right]$, 
a procedure also used in previous work~\cite{Mravlje2011,Zingl2019}.

\subsection{Comparison to Cuprate}

\begin{table}[h]
\begin{tabular}{|l|l|l|l|l|l|}
\hline
          & $d^7$ & LS $d^8$ & HS $d^8$ & $d^9$ & $d^{10}$ \\ \hline
NdNiO$_2$ & 0.05  & 0.12     & 0.28     & 0.50  & 0.04     \\ \hline
CaCuO$_2$ & 0.00  & 0.03     & 0.04     & 0.54  & 0.38     \\ \hline
\end{tabular}
\caption{Multiplet occurrence probabilities for the \NNO{} and \CCO{} Ni/Cu-$d$ shells obtained from one-shot DFT+DMFT calculations.}
\label{tab:cuprate_occ_prob}
\end{table}

To investigate the effect of using projectors on \CCO{}, we run one-shot DFT+DMFT calculations on \NNO{} and \CCO{} using a wide energy window of $\SI{-10}{eV}$ to $\SI{10}{eV}$ and use a 5 $d$ orbital impurity model. We run both calculations on the stoichiometric (nominal $d^9$) materials. 

Table \ref{tab:cuprate_occ_prob} shows the resulting multiplet occurrence probabilities. Similar to the Wannier function case \cite{karp2020manybody}, the cuprate has a high percentage of $d^{10}$ and a relatively small amount of $d^8$.

\subsection{Double Counting}
We examine the effect of changing the double counting correction by running a one-shot DFT+DMFT calculation on stoichiometric \NNO{} using the around mean field (AMF) double counting scheme and compare the results to those of the one-shot FLL scheme. 

\begin{table}[h]
\begin{tabular}{|l|l|l|l|l|l|}
\hline
    & $d^7$ & LS $d^8$ & HS $d^8$ & $d^9$ & $d^{10}$ \\ \hline
FLL & 0.05  & 0.12     & 0.28     & 0.50  & 0.04     \\ \hline
AMF & 0.10  & 0.15     & 0.39     & 0.34  & 0.02     \\ \hline
\end{tabular}
\caption{Comparison of multiplet occurrence probabilities for the \NNO{} Ni-$d$ shell using FLL and AMF double counting schemes.}
\label{tab:amf_occ_prob}
\end{table}

\begin{table}[h]
\begin{tabular}{|l|l|l|l|l|l|}
\hline
    & $d_{z^2}$ & $d_{x^2-y^2}$ & $d_{xy}$ & $d_{xz/yx}$ & total \\ \hline
FLL & 1.61      & 1.12          & 1.96     & 1.92        & 8.52  \\ \hline
AMF & 1.46      & 1.05          & 1.96     & 1.90        & 8.27  \\ \hline
DFT & 1.58      & 1.19          & 1.95     & 1.89        & 8.50  \\ \hline
\end{tabular}
\caption{Orbital occupancies obtained using FLL and AMF double counting schemes compared with the corresponding DFT values.}
\label{tab:amf_orb_occ}
\end{table}

Table \ref{tab:amf_occ_prob} shows the resulting multiplet occurrence probabilities and Table \ref{tab:amf_orb_occ} shows the orbital occupancies. The AMF double counting scheme empties out the $e_g$ orbitals, particularly the $d_{z^2}$ orbital, more than the FLL scheme. Consequently, the AMF results in significantly more high spin $d^8$ than FLL.

\section{Fat band analysis}
\label{app:fat_bands}

\begin{figure*}[]
    \centering
    \includegraphics[width = \linewidth]{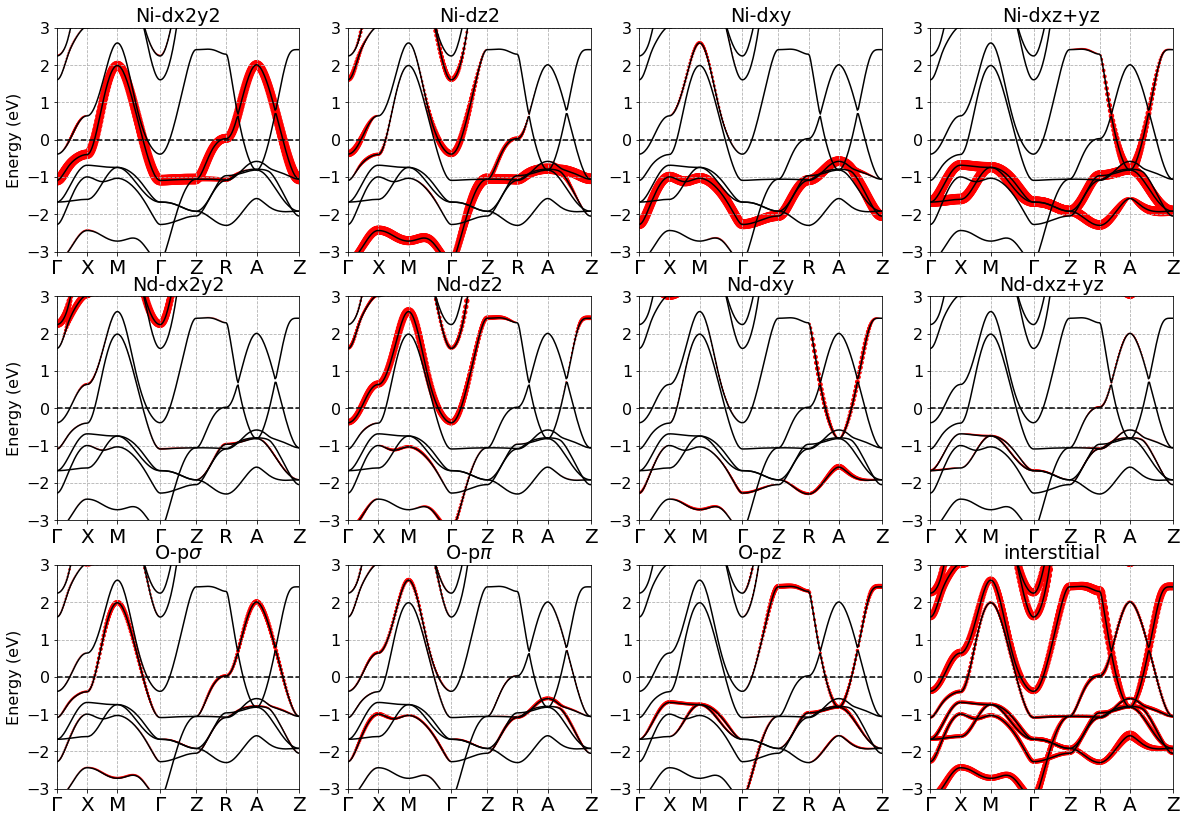}
    \caption{Orbital character of energy bands for undoped (nominal $d^9$) \NNO{}}
    \label{fig:NNO_fatbands}
\end{figure*}

\begin{figure*}
    \centering
    \includegraphics[width = .9\linewidth]{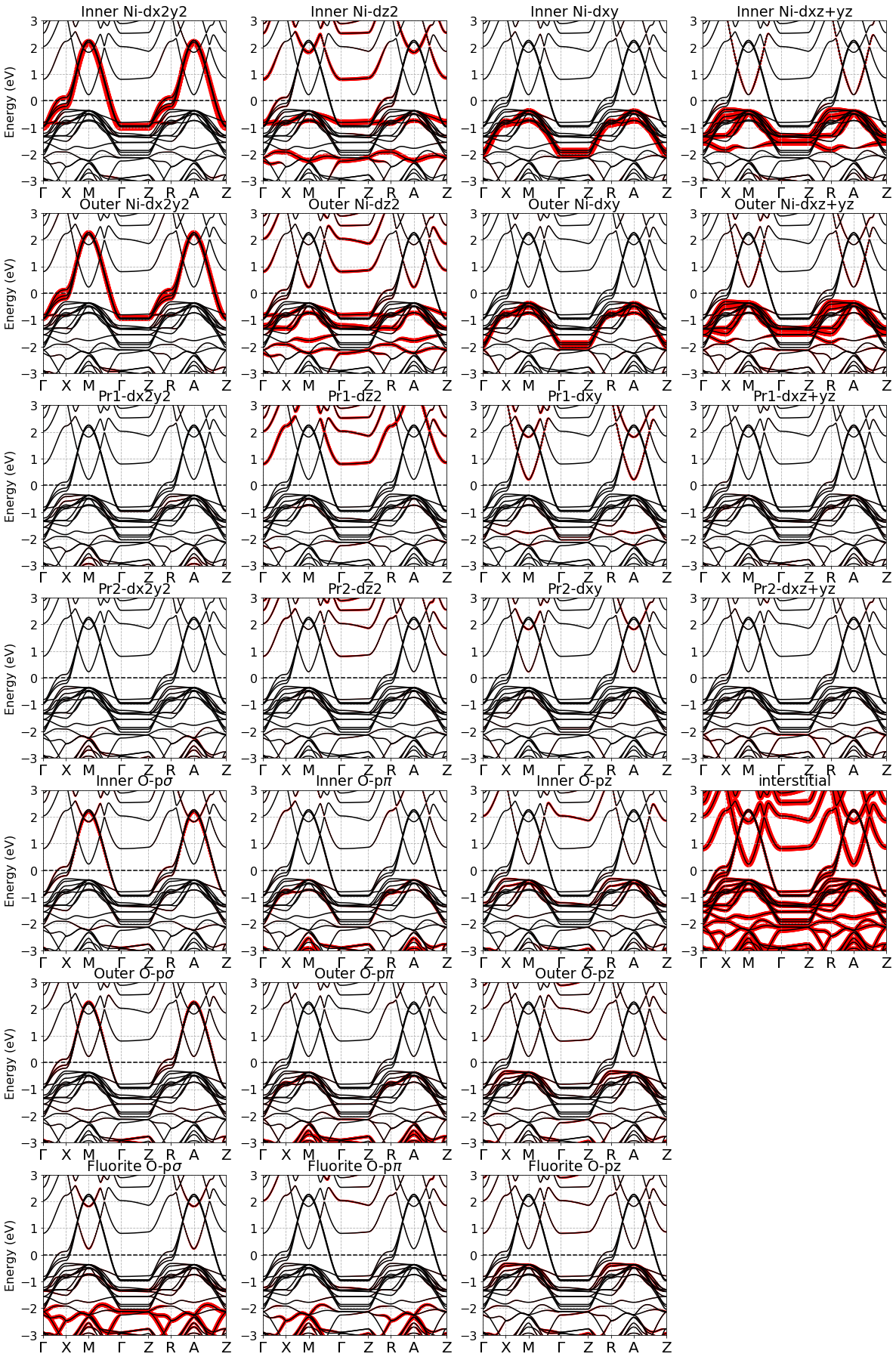}
    \caption{Orbital character of energy bands for undoped (nominal $d^{8.67}$) \PNO{}}
    \label{fig:PNO_fatbands}
\end{figure*}

Figs.~\ref{fig:NNO_fatbands} and \ref{fig:PNO_fatbands} show the orbital character of the near Fermi energy region for stoichiometric \NNO{} and \PNO{}, respectively. The plots show that the main bands crossing the Fermi energy are of Ni-$d_{x^2-y^2}$ character with some O-$p_\sigma$ admixture. For \NNO{}, Fig.~\ref{fig:NNO_fatbands} shows that the $\Gamma$ pocket is a mixture of Ni-$d_{z^2}$ and Nd-$d_{z^2}$, and the $A$ pocket is a mixture of Ni-$d_{xz/yz}$ and Nd-$d_{xy}$. For the case of stoichiometric \PNO{}, Fig.~ \ref{fig:PNO_fatbands} shows that the band which goes down to $\sim \SI{0.8}{eV}$ at the $\Gamma$ point, and goes below the Fermi energy upon electron doping, is a mixture of Ni-$d_{z^2}$ and Pr1-$d_{z^2}$, but does not contain significant amounts of Pr2. Likewise, the band which goes down to $\sim \SI{0.2}{eV}$ at the $M$ point and goes below the Fermi level upon doping is of Ni-$d_{xz/yz}$ and Pr1-$d_{xy}$ character, but not any significant Pr2 character. 

\onecolumngrid
\twocolumngrid

\clearpage
\bibliography{references.bib}

\end{document}